\documentclass[12pt]{article}
\usepackage{graphicx,latexsym}

  \def\pd{\partial} \def\pp{\prime}  \def\b{\beta} \def\dl{\delta} \def\s{\sigma}   \def\eps{\epsilon} 
 \def\lam{\lambda} \def\Lam{\Lambda} \def\gm{\gamma} \def\Gm{\Gamma}
 \def\Om{\Omega} \def\nb{\nabla} \def\sq{\sqrt} \def\e{\hbox{\large \it e}}
\def\half{\frac{1}{2}} \def\fr{\frac}  

\def\P{{\rm P}} \def\QG{{\rm QG}}  \def\M{{\rm M}}

\def\hphi{{\hat \phi}}

\def\bnb{{\bar \nabla}} \def\bg{{\bar g}} \def\bDelta{{\bar \Delta}} \def\bR{{\bar R}}

  \def\bx{{\bf x}}   
      \def\D{{\bf D}}

\def\lap3{~| \!\!\! \partial^2} \def\dlap3{~| \!\!\! \partial^4}

\begin{document}

\begin{center}
{\Large {\bf Quantum Gravity Effective Action Provides Entropy of The Universe}} \\ 
\end{center}

\begin{center}
{\sc Ken-ji Hamada}
\end{center}

\begin{center}
{\it Institute of Particle and Nuclear Studies, KEK, Tsukuba 305-0801, Japan \\ and Graduate Institute for Advanced Studies, SOKENDAI, Tsukuba 305-0801, Japan}
\end{center}

\begin{abstract}
\noindent
The effective action in renormalizable quantum theory of gravity provides entropy because the total Hamiltonian vanishes. Since it is a renormalization group invariant that is constant in the process of cosmic evolution, we can show conservation of entropy, which is an ansatz in the standard cosmology. Here, we study  renormalizable quantum gravity that exhibits conformal dominance at high energy beyond the Planck scale. The current entropy of the universe is derived by calculating the effective action under the scenario of quantum gravity inflation caused by its dynamics. We then argue that ghost modes must be unphysical but are necessary for the Hamiltonian to vanish and for entropy to exist in gravitational systems. 
\end{abstract}

\section{Introduction}

The origin of a huge amount of cosmic entropy, most of which is currently carried by the cosmic microwave background radiation (CMB), remains a big mystery. Within the framework of Einstein's theory of gravity, it cannot be satisfactorily explained without introducing an unknown scalar field as a source of all matter. Moreover, the standard cosmology assumes that it is conserved during evolution \cite{kt}. The conservation law has been explained thermodynamically after the big bang, but we wonder if that is really sufficient. The observed CMB anisotropy spectra \cite{wmap13, planck} show that the whole universe is entangled in every corner. It suggests that there was a moment in the past when they were correlated with each other. Therefore, it is natural to think that the origin of entropy and its conservation law were ruled by the laws of physics at that time. In this paper, we argue that it is derived from quantum spacetime states described by renormalizable quantum gravity.

In general, the effective action in renormalizable quantum field theory is finite and is renormalization group (RG) invariant \cite{collins}. If the theory is diffeomorphism invariant, the energy-momentum tensor is also finite, i.e. a normal product, and RG invariant \cite{acd, bc, hathrellS, hathrellQED, hamada14GC}. Furthermore, when gravity is quantized, the whole energy-momentum tensor vanishes.

Although this fact is widely known as the Hamiltonian and momentum constraints \cite{adm, dewitt, bhs}, it is best to use the Schwinger-Dyson (SD) equation to see that this holds as an identity at the quantum level \cite{hamada22}. Considering the partition function defined by the path integral over the gravitational field, $e^{i\Gm_\QG} = \int [dg] e^{iI}$, where $I$ is a renormalizable gravitational action and $\Gm_\QG$ is the effective action, then it can be expressed as
\begin{eqnarray}
     i \half \langle \sq{-g} \, T^{\mu\nu} (x) \rangle = \int [dg] \fr{\dl}{\dl g_{\mu\nu}(x)} e^{iI} = 0 .
       \label{SD equation}
\end{eqnarray}
Strictly speaking, we need to define the path integral measure as shown later; thus it is still a symbolic formula at this point, but captures the essence.

The partition function of quantum gravity represents sum over states of quantum spacetime. Since the effective action is defined by the logarithm of the partition function, it is entropy itself if the total Hamiltonian of the system vanishes.\footnote{%%%%%(footnote)%%%
It may be helpful to recall that effective action corresponds to free energy in ordinary quantum field theory. However, note that there are no ordinary particle states in quantum gravity considered here; thus, no concept of temperature or thermal equilibrium, but the formula is defined as purely counting the number of states. The inability to define temperature and the vanishing of the Hamiltonian are thus complementary.} %%%%%%%
Thus, the quantum gravity effective action provides a statistical entropy for the state of the universe. That is,
\begin{eqnarray}
       \Gm_\QG = -S_{\rm Univ} ,
          \label{entropy and effective action}
\end{eqnarray}
where $S_{\rm Univ}$ is the entropy, which is conserved as an RG invariant. This is a general conclusion derived from diffeomorphism invariance and renormalizability.

On the other hand, as if to deny this idea, it is often emphasized that conventional renormalization methods are not applicable to quantum theory of gravity. In fact, Einstein's theory of gravity is unrenormalizable. To escape this problem, researchers usually introduce a finite ultraviolet (UV) cutoff in the Planck scale and often think of it as an entity of spacetime quantization. However, this idea breaks diffeomorphism invariance. To preserve diffeomorphism invariance, renormalizability is necessary: then spacetime exists continuously while fluctuating greatly even beyond the Planck scale.

The problem with renormalization is relevant to the existence of singularities. Since the Einstein-Hilbert action does not contain the Riemann tensor $R_{\mu\nu\lam\s}$ that controls curvature, singularities cannot be removed. Introducing positive-definite fourth-derivative actions involving $R_{\mu\nu\lam\s}^2$ make gravity renormalizable because coupling constants become dimensionless \cite{stelle, tomboulis, ft}, and furthermore, singularities are statistically forbidden as objects for which the action diverges. But if we formulate it simply in the usual graviton picture defined by a perturbation expansion around flat spacetime, another difficulty arises called the ghost problem.

However, the existence of ghost modes itself is not the problem. The local particle world has to be positive definite, but the universe as a whole is not so. Ghost modes are rather essential elements for the total Hamiltonian to vanish, i.e., to preserve diffeomorphism invariance. As a matter of fact, Einstein's theory of gravity has a ghost mode due to the indefinitness of the Einstein-Hilbert action, which allows non-trivial solutions such as the Friedmann solution to exist.

The ghost mode in renormalizable quantum gravity exists in a different sense from that in Einstein's theory of gravity. It exists as a characteristics of higher-derivative fields even though the action is positive definite. That is, the gravitational field itself does not become a ghost, but it is hidden within the field as a sub-mode. In any case, if all modes were positive definite, the only state in which the Hamiltonian vanishes would be the trivial vacuum, and thus, would be no entropy. The ghost mode causes problems only when it appears as a physical particle.

The particle picture propagating through a specific spacetime is not appropriate to describe spacetime that  itself fluctuates quantum mechanically. In order to solve the ghost problem, this picture should be discarded. Renormalizable quantum gravity that exhibits background freedom asymptotically has been proposed as a way to achieve this \cite{hs, hamada02, hamada14re, hm16, book}. A distinctive feature of the theory is that it implements a novel perturbation technique expanding around spacetime where the Weyl tensor $C_{\mu\nu\lam\s}$ vanishes instead of the conventional perturbation expansion around flat spacetime. In the UV limit of this perturbation, the conformal mode of the gravitational field determining distance is not subject to any restrictions, and thus fluctuates largely in a non-perturbative manner.\footnote{%%%%%(footnote)%%%
Its statistical behavior appears to be in the same universality class as 4-dimensional simplicial quantum gravity \cite{hey}.} %%%%%%% 
Hence, scalar fluctuations dominate in the early universe. The correctness of this approach is supported by the idea of inflation \cite{guth, sato, starobinsky, hy, hhy06, hhy10, hamada23}.

This theory has a new mechanism for constraining ghosts, called the BRST conformal invariance\footnote{%%%(footnote)%%%%
BRST is an acronym for Becchi, Rouet, Stora, and Tyutin, which is used to emphasize that this is a gauge symmetry.} %%%%%%%%
\cite{riegert, am, amm92, amm97, hh, hamada12M4, hamada12RS3}, which arises as part of diffeomorphism invariance in the UV limit \cite{hs, hamada02, hamada14re, hm16, book}. It represents the background freedom that all different conformally-flat spacetimes are gauge equivalent. The BRST conformal invariance condition shows that all ghost modes involved in the gravitational field are unphysical and can never be seen, while there are an infinite number of physical states, which are only scalar-type (primary scalars): there are no tensor-type states \cite{hh, hamada12M4, hamada12RS3}. For the unitarity issue and the definition of this symmetry, see Appendices A and B.

In this paper, we show that the effective action of this quantum gravity indeed provides the entropy of the present universe.

\section{Renormalizable Quantum Gravity} 

The action of renormalizable quantum gravity that indicates conformal dominance asymptotically is given by \cite{hs, hamada02, hamada14re, hm16, book}
\begin{eqnarray}
   I = \int d^4 x \hbox{$\sq{-g}$}  \biggl\{
      -\fr{1}{t^2} C_{\mu\nu\lam\s}^2 -b G_4 
       + \fr{1}{\hbar} \left( \fr{1}{16\pi G}R   + {\cal L}_\M \right) \biggr\} .
\end{eqnarray}
The first is the Weyl action, and the second is the Euler term, where $G_4=R^2_{\mu\nu\lam\s}-4R^2_{\mu\nu} + R^2$, both of which are conformally invariant. The third with the scalar curvature $R$ is the Einstein-Hilbert action, where $G$ is the Newton constant. The fourth ${\cal L}_\M$ denotes matter field actions that are conformally invariant in the UV limit, but their specific expressions are not necessary below.

The Planck constant $\hbar$ appears only before the lower derivative terms, which is due to the fact that the gravitational field is completely dimensionless. This implies that all of the fourth-derivative gravitational actions here and below describe purely quantum dynamics and produce entropy of spacetime. In the following, $\hbar=1$.

Using the conformal mode $\phi$ and the traceless tensor mode $h^\mu_{~\nu}$, the gravitational field is decomposed as 
\begin{eqnarray}
     g_{\mu\nu} = e^{2\phi} \bg_{ \mu\nu}
         \label{decomposition of metric field}
\end{eqnarray} 
with $\bg_{\mu\nu}= (\eta e^h )_{\mu\nu}=\eta_{\mu\lam} ( \dl^\lam_{~\nu } + h^\lam_{~\nu}+ h^\lam_{~\s} h^\s_{~\nu}/2 + \cdots)$. The conformal factor $e^{2\phi}$ is handled nonperturbatively, while $h^\mu_{~\nu}$ is expanded as being small at the UV limit. The coupling constant for the expansion is $t$ and is introduced in front of the Weyl action, while $b$ is not an independent coupling because the Euler term does not contain a kinetic term. The flat metric $\eta_{\mu\nu}=(-1,1,1,1)$ defines the comoving frame with coordinates $x^\mu =(\eta,\bx)$.

The key to quantization is to rewrite the theory into a quantum field theory defined on the familiar flat spacetime. The partition function is then expressed as \cite{hs, hamada02, hamada14re, hm16, book, hh, hamada12M4, hamada12RS3}
\begin{eqnarray}
    e^{i\Gm_\QG} = \int [dg]_g \, e^{iI(g)} = \int [d\phi dh]_\eta \,e^{i S(\phi,\bg)+i I(g)} ,
         \label{definition of path integral}
\end{eqnarray}
where $S$ is the Wess-Zumino action \cite{wz} for the conformal anomaly \cite{cd, ddi, duff, duff94}, which is necessary to preserve diffeomorphism invariance when rewriting the path integral measure to the practical measure on flat spacetime. The SD equation (\ref{SD equation}) is precisely defined for the fields $\phi$ and $h^\mu_{~\nu}$, which are nothing but the equations of motion for them and are derived from the effective action.\footnote{%%%%%(footnote)%%%%%
When quantizing gravity using dimensional regularization that preserves diffeomorphism invariance manifestly, the SD equation (\ref{SD equation}) holds at it is, including contributions from the measure, without rewriting it as in (\ref{definition of path integral}), because it is a regularization that does not depend on how to choose the measure \cite{hamada14re}.} %%%%%%%

The Wess-Zumino action $S$ is responsible for fourth-derivative dynamics of the conformal mode $\phi$. The Riegert action associated with the Euler-type conformal anomaly \cite{riegert} that remains even in the zeroth order of $t$ is particularly important and is
\begin{eqnarray}
  S_{\rm R} = \int d^4 x  \biggl\{ - \fr{b_c}{(4\pi)^2} B \biggl[ 2\phi \bDelta_4 \phi  
                   + \biggl( \bar{G}_4 - \fr{2}{3} \bnb^2 \bR  \biggr)\, \phi 
                          \biggr]  \biggr\} ,
\end{eqnarray}
which provides a kinetic term of $\phi$.\footnote{%%%%(footnote)%%%%
\label{note on higher order wz action}At higher orders of $t$, the Wess-Zumino actions such as $\phi^{n+1} (2 \bDelta_4 \phi + \bar{G}_4-2\bnb^2 \bR/3)$, $\phi^n \bar{C}_{\mu\nu\lam\s}^2$, and also $\phi^n {\bar F}_{\mu\nu}^2$ for a gauge field $F_{\mu\nu}$ with $n \geq1 $ arise. These terms are eventually incorporated into the running coupling constant \cite{hamada20CA}.} %%%%%% 
The quantities with the bar denote those defined by $\bg_{\mu\nu}$, and $\sq{-g} \Delta_4$ is a fourth-order differential operator  that is conformally invariant for scalars and is defined by $\Delta_4 = \nb^4 + 2R^{\mu\nu} \nb_\mu \nb_\nu - 2R \nb^2/3 + \nb^\mu R \nb_\mu/3$. The lowest coefficient is given by $b_c = ( N_{\rm X} + 11 N_{\rm W}/2 + 62 N_{\rm A} )/360 + 769/180$ \cite{ft, hs, amm97, cd, ddi, duff, duff94}, where $N_{\rm X}$, $N_{\rm W}$, and $N_{\rm A}$ are the number of scalar fields, Weyl fermions, and gauge fields in the matter sector, respectively, and $b_c = 7.0$ for the Standard Model. A correction by $t$ is denoted as $B = 1 - \gm_1 t^2/4\pi + o(t^4)$, where $\gm_1$ is positive.

The beta function is negative as $\mu dt/d\mu = -\b_0 \, t^3 + o(t^5)$ with $\b_0 = [ ( N_{\rm X} + 3 N_{\rm W}+ 12 N_{\rm A} )/240 + 197/60 ]/(4\pi)^2$ \cite{ft, hs, amm97, cd, ddi, duff, duff94}, where $\mu$ is an arbitrary mass scale introduced upon quantization. The dynamics of the tensor mode cause running of the coupling constant, expressed as $\bar{t}^2(Q)= [\b_0 \log (Q^2/\Lam_{\rm QG}^2)]^{ -1}$, where $Q^2=q^2/e^{2\phi}$ is physical momentum squared, $q^2$ is comoving momentum squared, and the $\phi$-dependence comes from the Wess-Zumino actions \cite{hamada02, hhy06, hamada20CA}. The effective action is then expressed in the form that $t^2$ is replaced with $\bar{t}^2(Q)$, such as $- [1/\bar{t}^2 (Q)] \sq{-g} \, C^2_{\mu\nu\lam\s}$ for the Weyl part.\footnote{%%%%(footnote)%%%
\label{note of weyl effective action}More precisely, this is a rewrite of  $-[1/t^2 - 2 \b_0 \phi + \b_0 \log (q^2/\mu^2) ] \sq{-g} \, C^2_{\mu\nu\lam\s}$, where the second is the Wess-Zumino action and the third is a loop correction. The inside of the square brackets is summarized to the form of $1/\bar{t}^2 (Q)$. This effective action form holds even when higher-order corrections are involved, in which case the running coupling constant is expressed by incorporating contributions from the higher-order Wess-Zumino actions listed in Footnote \ref{note on higher order wz action}.} %%%%%% 
The new dynamical energy scale $\Lam_{\rm QG} \, (=\mu \, e^{-1/2\b_0 t^2})$ is a physical constant, i.e., an RG invariant satisfying $d \Lam_{\rm QG}/d\mu=0$, as is the Planck mass \cite{hm17}. The correlation length is given by $\xi_\Lam =1/\Lam_\QG$, so that spacetime will be dynamically discretized by this length \cite{hamada20LME}.

The inflationary solution exists only for $M_\P > \Lam_\QG$, where $M_\P$ is the reduced Planck mass \cite{hy, hhy06, hhy10, hamada23}. This relation also serves as a unitarity condition to ensure that ghost modes do not appear as particles in the world after the spacetime phase transition. In quantum spacetime, all ghost modes are constrained by the BRST conformal invariance, but this cannot be applied after the transition. Therefore, the transition must occur below the Planck scale so that ghost gravitons with masses of about $M_\P$ are not created.

\section{Entropy of The Universe}

Most of entropy in the quantum gravity state is carried by the conformal mode rather than tensor or matter modes. Its dynamics cause inflation, and entropy in that period will give the entropy of the universe. The homogeneous component of the conformal mode is then responsible for it significantly; it is denoted by $\hphi$ and an average value of spacetime fluctuations.

The effective action for this mode, which plays the main part of inflation dynamics, is given by
\begin{eqnarray}
    \Gm_\QG = V_3 \int d\eta \biggl\{  - \fr{b_c}{8\pi^2} B \hphi \pd_\eta^4 \hphi 
                 + 3 M_\P^2 e^{2\hphi} \bigl( \pd_\eta^2 \hphi + \pd_\eta \hphi \pd_\eta \hphi \bigr) \biggr\} ,
           \label{effective action}
\end{eqnarray}
where $V_3= \int d^3 \bx$, and the first and second terms are contributions from the Riegert and Einstein-Hilbert sectors, respectively.

Since $C_{\mu\nu\lam\s} \simeq 0$ during most of the period of inflation, the contribution from the Weyl part of the effective action to entropy is sufficiently small and can be ignored. The main role of the Weyl sector is to transfer the entropy carried by the conformal mode to matter fields during the spacetime phase transition. The reason why the matter sector is disregarded is that the matter energy density is almost zero during the inflation period as stated below.

The equation of motion for the homogeneous mode is given by
\begin{eqnarray}
      - \fr{b_c}{4\pi^2} B \pd_\eta^4 \hphi  
          +  6 M_\P^2 e^{2\hphi} \bigl(  \pd_\eta^2 \hphi  + \pd_\eta \hphi  \pd_\eta \hphi  \bigr) = 0 .
        \label{homogeneous equation of motion}
\end{eqnarray}
This equation has a stable inflationary de Sitter solution at $t \to 0 \, ( B \to 1)$ \cite{hy}.\footnote{%%%%(footnote)%%%
It has been confirmed that the solution converges to (\ref{inflationary solution}) even if the initial conditions are changed significantly.} %%%% 
Introducing the proper time defined by $d\tau = a d\eta$ with the scale factor $a=e^\hphi$ and using the Hubble variable $H= \pd_\tau a/a$, the solution is expressed as
\begin{eqnarray}
       H = H_\D ,   \qquad    H_\D = \sq{\fr{8\pi^2}{b_c}} M_\P .
           \label{inflationary solution}
\end{eqnarray}
This shows that the scale factor increases exponentially as $a = e^{H_\D \tau}$, where let it be unity at $\tau =0$ representing infinite energy so that $V_3$ is the spatial volume far before inflation begins. Since $b_c$ is about $10$, $H_\D$ is also a Planck scale and is slightly larger than $M_\P$.

In addition, letting $\rho$ be an energy density of matter fields, we obtain an energy conservation equation from the vanishing of the time-time component of the energy-momentum tensor as \cite{hhy06}
\begin{eqnarray}
     \fr{b_c}{8\pi^2} B \bigl( 2 \pd_\eta^3 \hphi \pd_\eta \hphi -\pd_\eta^2 \hphi \pd_\eta^2 \hphi  \bigr) 
      - 3 M_\P^2 \e^{2\hphi} \pd_\eta \hphi \pd_\eta \hphi  + \e^{4\hphi} \rho = 0 .
         \label{energy conservation in conformal time}
\end{eqnarray}
Substituting the de Sitter solution (\ref{inflationary solution}) into this equation with $B=1$ yields $\rho=0$. Thus, matter is generated when the coupling constant $t$ increases so that $B$ decreases.

The inflationary expansion starts at the Planck scale $H_\D$ and terminates at the dynamical scale $\Lam_\QG$ where a spacetime phase transition occurs \cite{hy, hhy06, hhy10, hamada23}. The running coupling constant $\bar{t}^2$ indicates that spacetime is initially conformally invariant, while when energy drops to the vicinity of $\Lam_\QG$, $\bar{t}^2$ increases rapidly and deviates from such a spacetime. The phase transition is expressed as a process in which the conformal gravity dynamics disappear as $\bar{t}^2$ grows and the spacetime shifts to the present universe dominated by the Einstein-Hilbert action. At this time, since energy of the whole system is preserved to be zero as in (\ref{energy conservation in conformal time}) due to diffeomorphism invariance, the disappeared energy of quantum gravity is transferred to $\rho$, causing the big bang.\footnote{%%%%%%(footnote)%%%
Interactions that cause the big bang are given by the Wess-Zumino actions, such as $\phi \sq{-g} \, F^2_{\mu\nu}$ and $\phi \sq{-g}\, C^2_{\mu\nu\lam\s}$. These interactions are open near the phase transition as the running coupling constant increases.} %%%%%%
Entropy of the universe (\ref{entropy and effective action}) is also released to matter while preserving its total amount.

To describe such dynamics, we have proposed a model that approximates the running coupling constant by a time-dependent mean field $\bar{t}^2(\tau)= [\b_0 \log (\tau_ \Lam^2/\tau^2)]^{-1}$ that diverges at the dynamical time $\tau_\Lam=1/\Lam_\QG$ \cite{hhy06, hhy10, hamada23}. The running behavior is incorporated by replacing $t^2$ with $\bar{t}^2(\tau)$, and the replaced $B$ is denoted as $\bar{B}(\tau)$. The disappearance of conformal gravity dynamics is expressed by assuming its form as $\bar{B} (\tau) =[1+\gm_1 {\bar t}^2(\tau)/4\pi]^{-1}$, where $\b_0$ and $\gm_1$ are treated as phenomenological parameters. However, to be able to carry out calculations analytically, here we mainly consider a radically simplified model so that $\bar{t}^2$ remains almost zero and diverges sharply at $\tau_\Lam$, so that $\bar{B}$ is almost $1$ and abruptly goes to zero at $\tau_\Lam$ as a step function. In the meanwhile, we refer to the original unsimplified dynamical model as appropriate.

\begin{figure}[h]
\begin{center}
\includegraphics[scale=0.32]{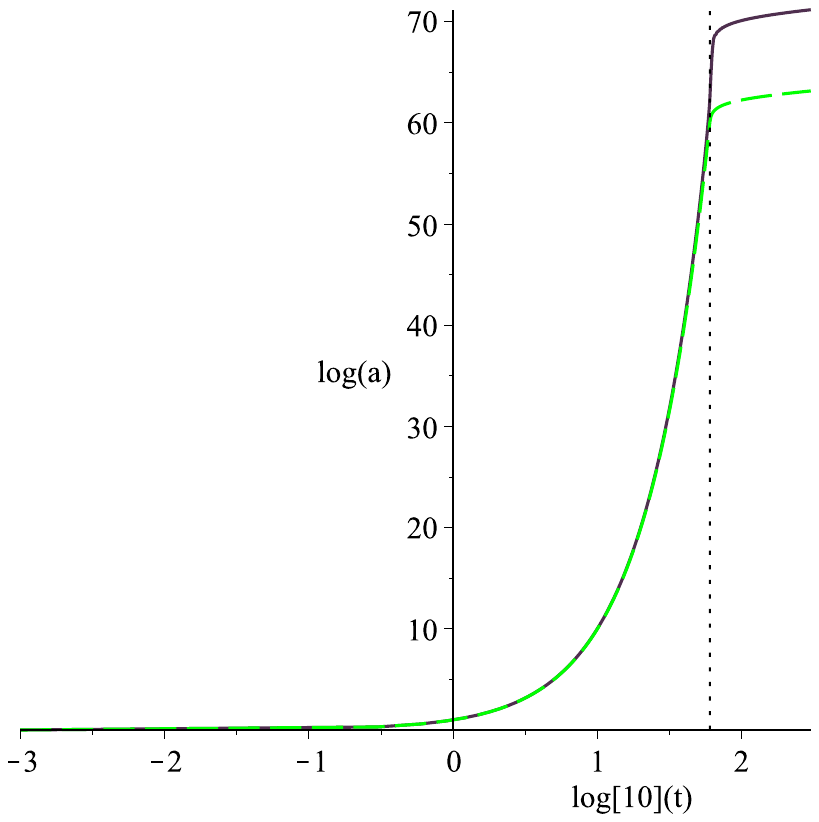}
\includegraphics[scale=0.32]{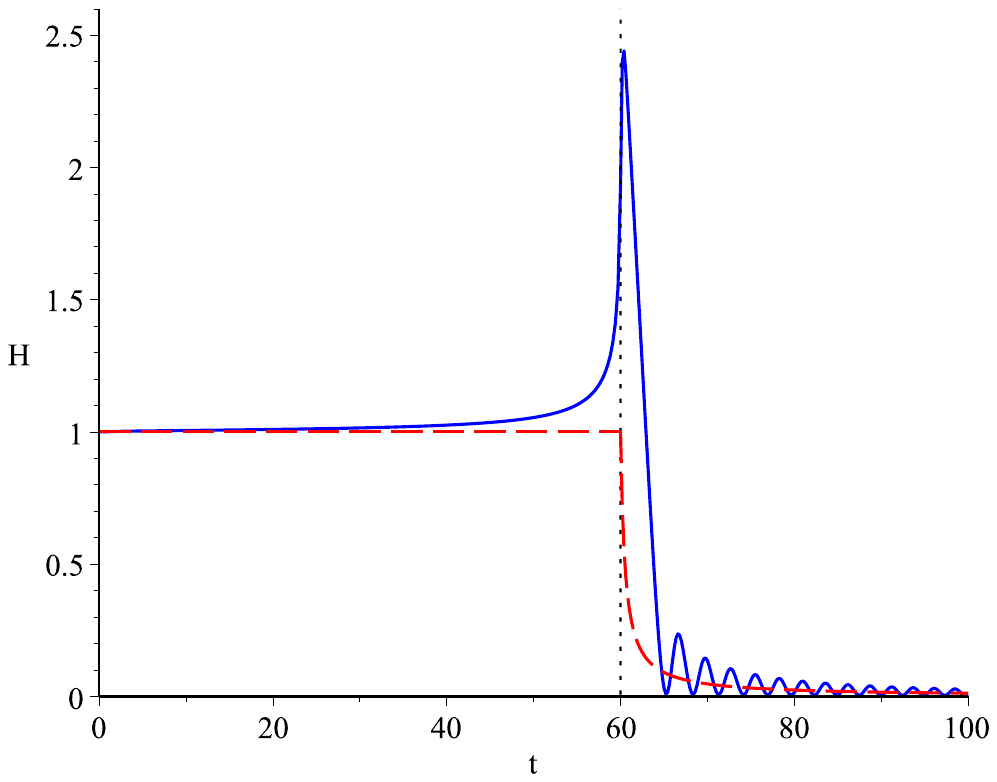}
\end{center}
\vspace{-4.7cm}
\caption{\label{plot of inflation scenario}{\small The inflationary solution presented by renormalizable and asymptotically background-free quantum gravity: The left is the time evolution of the scale factor $a$, and the right is that of the Hubble variable $H$, where $H_\D$ is normalized to unity, and $t \,(=H_\D \tau)$ is a normalized proper time. The solid lines (violet and blue) are the solutions of the dynamical model approximating the running coupling constant with the time-dependent mean field, and the dashed lines (green and red) are the radically simplified case. The vertical dotted line represents the dynamical time $\tau_\Lam$, which is set to $60$ and at which the spacetime phase transition occurs. After time bigger than $\tau_\Lam$, the variables are calculated using the low-energy effective gravity theory defined by derivative expansions centered on the Einstein-Hilbert action (see \cite{hhy06} in detail), while the dashed lines display the Friedmann solution.}}
\end{figure}

The number of e-foldings ${\cal N}_e = \log [a(\tau_\Lam)/a(\tau_\P)]$ from the Planck time $\tau_\P =1/H_\D$ to the dynamical time $\tau_\Lam$ is then given by the ratio of two energy scales as
\begin{eqnarray}
        N = \fr{H_\D}{\Lam_\QG} .
\end{eqnarray}
If we solve (\ref{homogeneous equation of motion}) with the dynamical factor $\bar{B}(\tau)$ rather than the step function, $H$ becomes larger than $H_\D$ near the phase transition and the number of e-foldings becomes slightly larger than $N$ (see Fig.\ref{plot of inflation scenario}).

\begin{figure}[h]
\begin{center}
\includegraphics[scale=0.3]{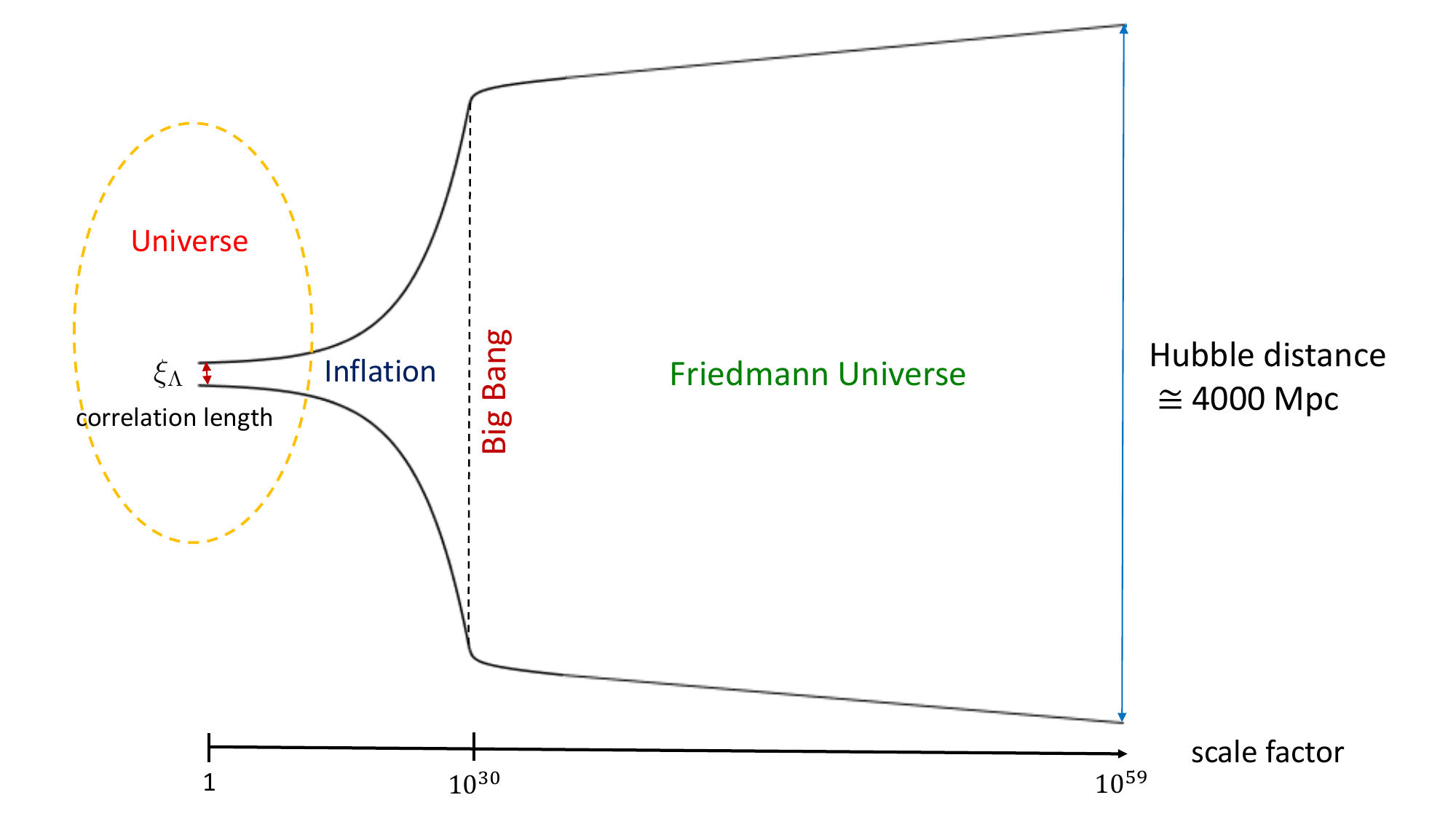}
\end{center}
\vspace{-5mm}
\caption{\label{sketch of cosmic evolution}{\small The evolution scenario of quantum gravity inflation: This shows that most of the universe we see today, represented by the Hubble distance, was originally the size of a quantum gravity excitation given by the correlation length $\xi_\Lam \, (=1/\Lam_\QG)$. That is, $1/H_0 \simeq 10^{59} \xi_\Lam$ or $H_0 \simeq \lam$. This suggests that the universe is an entangled quantum system.}}
\end{figure}

If $N$ is fixed, the number of e-foldings is determined, and the energy scale $\Lam_\QG$ is also determined. Scalar spacetime fluctuations, which account for most of initial fluctuations, have been shown to reduce in amplitude during inflation \cite{hhy06, hhy10, hamada23}. The magnitude at the phase transition can be roughly estimated as $\dl R/R \sim \Lam_\QG^2/12 H_\D^2 = 1/12N^2$, where the denominator is the de Sitter curvature. It should be the order of the CMB anisotropy, $10^{-4} \sim 10^{-5}$ \cite{wmap13, planck}. Furthermore, the comoving dynamical scale $\lam = \Lam_\QG/a_0$, where $a_0$ is the current scale factor, is approximately given by the current Hubble constant $H_0 \simeq 0.00023 \, {\rm Mpc}^{-1}$ to explain the sharp falloff in low multipole components of the CMB anisotropy spectrum \cite{hy}. From these conditions, $N$ is determined to be about $60$, and $\Lam_\QG$ is given on the order of $10^{17}$GeV, which is two orders of magnitude lower than the Planck energy \cite{hy, hhy06, hhy10, hamada23}. The magnitude of $a_0$ that expresses how much the universe has expanded since before inflation is determined to be about $10^{59}$ (see Fig.\ref{sketch of cosmic evolution}). From the previous studies, we have found that $\Lam_\QG$ is a tight parameter whose value cannot be changed very much.

Rewriting the effective action (\ref{effective action}) with the equation of motion (\ref{homogeneous equation of motion}) yields
\begin{eqnarray}
      \Gm_\QG &=& - 3 M_\P^2 V_3 \int d\eta \bigl( \hphi - 1 \bigr) e^{2\hphi} 
                         \bigl(  \pd_\eta^2 \hphi  + \pd_\eta \hphi  \pd_\eta \hphi  \bigr) .
            \label{inflation effective action}          
\end{eqnarray}
If the inflationary period given by (\ref{inflationary solution}) continues until it terminates at $\tau_\Lam$ in the proper time, the entropy of quantum gravity is given from the formula (\ref {entropy and effective action}) as
\begin{eqnarray}
      S_{\rm Univ} &=&  6 M_\P^2 H_\D^2 V_3 \int^{\tau_\Lam}_0 d\tau 
                     e^{3H_\D \tau} \bigl( H_\D \tau - 1 \bigr)
                            \nonumber \\
             &=& 2 M_\P^2 H_\D V_3 \biggl[ e^{3H_\D \tau_\Lam}  
                                \biggl( H_\D \tau_\Lam - \fr{4}{3} \biggr)   + \fr{4}{3} \biggr]   .            
\end{eqnarray}
Omitting the negligible last term and rewriting with the scale ratio $N$, we finally get
\begin{eqnarray}
      S_{\rm Univ} = 2 \sq{\fr{8\pi^2}{b_c}} M_\P^3 e^{3N} \biggl( N - \fr{4}{3} \biggr) V_3 .
           \label{entropy of universe}
\end{eqnarray}

The scalar curvature changes to zero during the spacetime phase transition, and matter is produced with energy density $\rho(\tau_\Lam) = 3 M_\P^2 H^2(\tau_\Lam)$. The universe then evolves according to the Friedmann solution. Since it a solution in which the Einstein-Hilbert action vanishes, spacetime itself has no entropy; thus, matters propagating through the spacetime, including gravitons, hold most of the entropy. Changing to this phase, we can define what is called thermal state.

The quantum gravity entropy (\ref{entropy of universe}) is inherited in matters as a conserved quantity. It should match the current entropy, which is given in density by \cite{kt},
\begin{eqnarray}
     s = \fr{S_{\rm Univ}}{V_3^{\rm today}} = \fr{2\pi^2}{45} g_{*S} T^3 \simeq 2.91 \times 10^3 \, {\rm cm}^{-3} ,
        \label{present entropy}
\end{eqnarray}
where $g_{*S}=3.91$ and $T= 2.73 \,{\rm K}=11.9{\rm cm}^{-1}$. When comparing the two expressions, note that $V_3$ in (\ref{entropy of universe}) is the initial three-dimensional volume before inflation. The universe has undergone inflation and the subsequent Friedmann eras, increasing the current scale factor to $a_0$. Therefore, from $V_3^{\rm today}= a_0^3 V_3$, the current entropy density can be expressed as
\begin{eqnarray}
      s = 2 \sq{\fr{8\pi^2}{b_c}} M_\P^3 e^{3N} \biggl( N - \fr{4}{3} \biggr) a_0^{-3} .
\end{eqnarray}
Here, fixing the parameters to $b_c=7$ and $a_0=10^{59}$ as derived before and finding the value of $N$ so that this agrees with (\ref{present entropy}), we get $N=62.2$, where $M_\P=2.436 \times 10^{18} \, {\rm GeV} = 1.235 \times 10^{32} \, {\rm cm}^{-1}$ is used. Thus, we find that the current entropy of the universe is consistent with the scenario of quantum gravity inflation.

The ratio between $2.73$K and $\Lam_\QG$ suggests that the universe expands about $10^{29}$ times after settling into the Friedmann spacetime. Therefore, the universe has to expand about $10^{30}$ times during the inflationary era. This corresponds to ${\cal N}_e \simeq 70$, which is greater than $60$. This contradiction can be resolved by considering a more realistic dynamical model described by the time-dependent running coupling constant above. Fig.\ref{plot of inflation scenario} shows an inflationary solution calculated by choosing the phenomenological parameters to be $\b_0=0.171$ and $\gm_1=0.1$, with $N=60$ so that entropy obtained by evaluating (\ref{inflation effective action}) with this solution numerically agrees with (\ref{present entropy}).\footnote{%%%%%(footnote)%%%
The calculation is done by piecewise integration in the proper time with the division width $\eps=10^{-4}$. At this time, to avoid numerical singularities, the initial time is set to $10^{-3}$ and the transition time is to $N-\eps$; then $\eps$ is determined so that the result does not change even if it is made smaller.} %%%%%%% 
In any case, quantum gravity can generate sufficient entropy.

\section{Conclusion}

The current entropy of the universe is derived from the effective action of renormalizable quantum gravity with asymptotic background freedom. The result suggests that entropy of the universe is originated from quantum gravity. The conservation of entropy can be shown from diffeomorphism invariance and renormalizability. Hence, these conditions should be taken as guiding principles even in the trans-Planckian world.

The ghost modes in gravity as unphysical beings are necessary to provide entropy while preserving the total Hamiltonian to be zero, and the zero-point energy then disappears due to them \cite{hamada22}. The cosmic time, which is a change in central value of the conformal-mode fluctuation given by the inflation and Friedmann solutions, is also one of things dynamically generated due to the existence of the ghost modes.

They also contribute to the structure formation of the universe. The ghost mode that makes the Einstein-Hilbert action unbounded below is responsible for the instability of the Friedmann universe: that is, the growth of fluctuations after the big bang resulting in the large-scale structure of the current universe. Conversely, spacetime fluctuations reduce in amplitude during the inflation period, giving the initial condition of the current universe \cite{hhy06, hhy10, hamada23}. This stability is due to the fact that the fourth-derivative gravitational action is positive definite even though it contains ghosts as sub-modes. In this way, these various ghost modes play a decisive role when considering the universe as a whole. We should pay more attention to this fact.

Instead of denying the existence of the ghost, we should consider what it brings to a quantum system of the entire universe.\footnote{%%%%(footnote)%%%
There is an idea to introduce boundaries and remove ghost modes by appropriately imposing boundary conditions \cite{maldacena}, but here we are considering a universe in which there are no boundaries leading to the unknown outside world, which is also a condition for the SD equation (\ref{SD equation}) to hold.} %%%%% 
For gravitational systems, since the Hamiltonian vanishes identically, we cannot directly apply no-go theorems proving that energy spectrum is unbound from below, such as the Ostrogradsky theorem \cite{ostrogradsky}.

Reconciling quantum mechanics and gravity requires a deep understanding of what it means for the Hamiltonian to vanish. The ghost mode reminds us of the Bohm's ``hidden variables'' in the sense that it is ``invisible, but being''. Note, however, that there is no such variable in special relativity, and the ghosts are necessary only when gravity has a substantial contribution.

The weak-field (graviton) approximation is nothing more than reducing the gravitational system to a system in special relativity. It is an ordinary quantum-mechanical system, in which case we consider the Hamiltonian eigenstate and no longer care about the vanishing of the Hamiltonian. All states in this system are subject to observation, thus ghost modes must be erasable. This approximation only applies to local particle worlds that do not affect the entire spacetime structure, and is improper for the trans-Planckian world where spacetime itself fluctuates quantum mechanically.

\appendix

\section{Notes on Unitarity or Reality}

Quantum spacetime where quantum gravity is fully activated is described in terms of a conformal field theory so that the particle picture no longer holds true and the scattering matrix is not defined. In such cases, the unitarity issue becomes easier to understant if  we consider quantum field theory in Euclidean space by performing the Wick rotation so that the partition function is described from the perspective of statistical mechanics like $\int e^{-{\cal I}}$.

If the Euclidean action ${\cal I}$ is real and bounded below, the Boltzmann weight $e^{-{\cal I}}$ is positive and finite, and thus, the path integral is correctly defined. In this case, the reality of the field is not lost, and unitarity holds: that is, two-point functions are positive-definite and structure constants (three-point functions) are real. This also state that a singular configuration where ${\cal I}$ is positively divergent can be removed as an unphysical state that does not exist stochastically because the Boltzmann weight vanishes, as mentioned in Introduction.

On the other hand, if ${\cal I}$ is an indefinite action that can be negative infinite, the path integral will diverge, and thus the field reality will be sacrificed to regularize the divergence. As can be seen from this fact, the path integral of quantum gravity cannot be correctly defined by the indefinite Einstein-Hilbert action alone; thus, the square of curvature is required to ensure positive definiteness.

Why does the ghost problem occur? Ultimately, the root cause of the problem lies in not adopting a perturbation expansion method suitable for describing quantum spacetime. Usually, it is carried out by employing a picture where gravitons propagate in flat spacetime. Einstein's theory of gravity allows us to describe spacetime with this picture because ghosts can be removed locally. In fourth-derivative quantum gravity, however, this description poses the problems that the positive- and negative-metric modes behave as independent physical degrees of freedom asymptotically; thus, ill-defined correlation functions among the negative-metric modes have physical meaning.

Conversely, if we consider physical quantities using the gravitational field as a ``fundamental field'' without separating into positive- and negative-metric sub-modes, correlation functions among the fields are correctly defined because the field actions, the Riegert and Weyl actions, are positive definite. Therefore, the reality of the field is never lost. In this argument, it is extremely important that in Einstein's theory of gravity, $\phi$ itself becomes a ghost mode, whereas in fourth-derivative quantum gravity, ghost modes do not appear unless the fundamental fields $\phi$ and $h_{\mu\nu}$ are further expanded into sub-modes.

Hence, in order to define fourth-derivative quantum gravity, it can be seen that when dealing with physical quantities, there must be some kind of constraints that do not allow the separation of the field into positive- and negative-metric modes so that the Hamiltonian vanishes. The BRST conformal invariance plays exactly this role.

\section{BRST Conformal Invariance}

The key to the asymptotically background-free quantum gravity lies in special properties of the core of perturbation theory. Unlike the usual weak-field expansion, free particle states do not appear as asymptotic states, as stated repeatedly. The core part is described by a special conformal field theory, and the coupling constant $t$ represents the degree of deviation from it.

In normal conformal transformations, when the coordinates are transformed to $x^\mu \to x^{\pp \mu}$, the line element $ds$ representing distance changes only by a conformal factor like $ds^2 \to ds^{\pp 2} = \Om^2 ds^2$. At this time, the metric field remains fixed. Conversely, it can also be defined as a transformation (Weyl transformation) that expands the metric field by a conformal factor without changing the coordinates.

In contrast, since the BRST conformal transformation is a manifestation of diffeomorphism in the UV limit, the metric field is also transformed under the coordinate transformation so that the line element remains unchanged as $ds^2 = ds^{\pp 2}$. Moreover, physical quantities must be conformally invariant in the BRST conformal field theory, which is very different from normal conformal field theory where only the vacuum is conformally invariant. This symmetry was first studied in two-dimensional quantum gravity \cite{polyakov, kpz, dk, david} and then extended to four dimensions.

Hence, normal conformal invariance refers to the theory becoming independent of a particular scale, although there is a metric that defines distance. On the other hand, conformal invariance in quantum gravity occurs when the distance itself fluctuates; thus, it truly represents a world without scale.

Specifically, it is expressed as follows \cite{hh, hamada12M4, hamada12RS3}. First of all, diffeomorphism is a transformation in which the line element $ds^2 = g_{\mu\nu} dx^\mu dx^\nu$ remains unchanged under the coordinate transformation $x^\mu \to x^{\pp \mu} = x^\mu - \xi^\mu$. If the gauge parameter $\xi^\mu$ is infinitesimal, then diffeomorphism is expressed as $\dl_\xi g_{\mu\nu}=g_{\mu\lam}\nb_\nu \xi^\lam + g_{\nu\lam}\nb_\mu \xi^\lam$. When the metric field is decomposed as in (\ref{decomposition of metric field}), diffeomorphism is expressed as
\begin{eqnarray}
   \dl_\xi \phi  &=&
          \xi^\lam \pd_\lam \phi   + \fr{1}{4} \pd_\lam \xi^\lam ,     
               \nonumber \\
     \dl_\xi h_{\mu\nu}  
     &=&  \pd_\mu \xi_\nu  +  \pd_\nu \xi_\mu  -  \fr{1}{2} \eta_{\mu\nu} \pd_\lam \xi^\lam  
           +  \xi^\lam \pd_\lam h_{\mu\nu}      
           +  \fr{1}{2} h_{\mu\lam}  \bigl(   \pd_\nu \xi^\lam     -  \pd^\lam \xi_\nu   \bigr) 
                 \nonumber \\
     && 
           +  \fr{1}{2} h_{\nu\lam}  \bigl(   \pd_\mu \xi^\lam     -  \pd^\lam \xi_\mu   \bigr) 
           + o(h^2)  ,
         \label{diffeomorphism for traceless tensor field}
\end{eqnarray}
where $\xi_\mu =\eta_{\mu\nu}\xi^\nu$.

The lowest term $\dl_\xi h_{\mu\nu} = \pd_\mu \xi_\nu + \pd_\nu \xi_\mu - \eta_{\mu\nu} \pd_\lam \xi^\lam/2$ in (\ref{diffeomorphism for traceless tensor field}) is the part that contributes the most in the UV limit. The kinetic term of the Weyl action given by the quadratic term of the traceless tensor field becomes invariant under this transformation. This gauge degree of freedom is fixed using the standard gauge-fixing method in gauge theory.

Even if these gauge degrees of freedom $\xi^\mu$ are fixed, there still remain $15$ gauge degrees of freedom $\zeta^\mu$ that satisfy the conformal Killing equation:
\begin{eqnarray}
      \pd_\mu \zeta_\nu + \pd_\nu \zeta_\mu 
                 - \fr{1}{2} \eta_{\mu\nu} \pd_\lam \zeta^\lam =0 .
           \label{conformal Killing equation}
\end{eqnarray}
For this gauge degree of freedom, since the lowest term in transformation law (\ref{diffeomorphism for traceless tensor field}) disappears and the next term becomes effective, diffeomorphism is expressed in conformal transformations as 
\begin{eqnarray}
    \dl_\zeta \phi  &=&
         \zeta^\lam \pd_\lam \phi + \fr{1}{4} \pd_\lam \zeta^\lam ,
              \nonumber \\
    \dl_\zeta h_{\mu\nu}  &=&
         \zeta^\lam \pd_\lam h_{\mu\nu} 
         + \half h_{\mu\lam} \bigl( \pd_\nu \zeta^\lam - \pd^\lam \zeta_\nu \bigr)
         + \half h_{\nu\lam} \bigl( \pd_\mu \zeta^\lam - \pd^\lam \zeta_\mu \bigr) .
             \label{conformal transformation}
\end{eqnarray}
where the field-independent shift term in the first, which is not present in normal conformal transformation, shows that this transformation is derived from diffeomorphism. For formulation in the BRST formalism, see \cite{hamada12M4, hamada12RS3}.

The BRST conformal invariance is thus a hidden symmetry that becomes visible when the theory is rewritten as a quantum field theory (\ref{definition of path integral}) in a background spacetime. It represents the background freedom that all theories on different backgrounds connected by the conformal transformation are gauge equivalent.

Physical quantities of quantum gravity in the high energy limit must be invariant under the conformal transformation (\ref{conformal transformation}). One of the characteristics of this transformation is that the right-hand side depends on the field. In an ordinary gauge transformation, the lowest term depends only on gauge parameter and is independent of the field. Under such a transformation, modes that make up the field become independent without mixing. On the other hand, under the transformation (\ref{conformal transformation}), the modes mix with each other and do not be gauge-invariant independently.

Therefore, gravitational ghost modes do not themselves become physical states. Physical states can be expressed as states in which the whole Hamiltonian exactly vanishes. The reason why they can exist as non-trivial states with entropy rather than an empty vacuum is because of the ghost mode. The BRST conformal invariance truly defines physical states that appears in the UV limit far beyond the Planck scale, and they are given by scalar states only.

%%%%%%%%%%%%%%%%%%%%%%
%%%   References   %%%
%%%%%%%%%%%%%%%%%%%%%%

\end{document}